**B-site substitutions and phase transitions in solid solutions on the base of $(Na_{0.5}Bi_{0.5})TiO_3$**

V. M. Ishchuk[1], L. G. Gusakova[1], N. G. Kisel'[1], D. V. Kuzenko[1], N. A. Spiridonov[1] and V. L. Sobolev[2]

[1]Science & Technology Center "Reactivelectron" of the National Academy of Sciences of Ukraine, Donetsk, 83049, Ukraine
[2]Deaprtment of Physics, South Dakota School of Mines and Technology, Rapid City, SD 57701

## Abstract

Substitutions of the Zr and Sn ions along with the $(In_{0.5}Nb_{0.5})$ and $(Fe_{0.5}Nb_{0.5})$ complexes for Ti in the $(Na_{0.5}Bi_{0.5})_{0.80}Ba_{0.20}](Ti_{1-y}B_y)O_3$ solid solutions impact the formation of phases in the process of the solid state synthesis and the relative stability of the antiferroelectric and ferroelectric phases.

Our studies show that the synthesis of solid solutions is a multi-step process during which a number of intermediate phases (depending on the solid solution composition and the annealing temperature) are formed. It is shown that the increase of the sintering temperature to 1000 – 1100 °*C* allows to obtain single phase solid solutions.

It is demonstrated that the increase of content of the substituting ion results in a linear variation in the size of the crystal lattice unit cell. At the same time it leads to a change in the relative stability of the antiferroelectric and ferroelectric phases according to the variation of the tolerance factor of the solid solution. Substitutions by ions with a radius larger than the radius of the initial ion leads to a decrease of the temperature of the ferroelectric-antiferroelectric phase transition whereas the substitutions of the smaller ions produce the opposite effect.

.

Lead zirconate titanate, lead zinc niobate-lead titanate, and lead magnesium niobate-lead titanate solid solutions have been the most popular for applications manifesting the largest values of piezoelectric parameters. Despite their outstanding piezoelectric properties, they are currently facing worldwide restrictions due to the toxicity of lead. There is an urgent need to develop lead-free materials with high piezoelectric parameters. The $(Bi_{1/2}Na_{1/2})TiO_3$ -based solid solutions are the object of intense studies at present. The most investigated among them are the $(1-x)(Bi_{1/2}Na_{1/2})TiO_3$ - $xBaTiO_3$ (BNT–BT) solid solutions.[1-4] Their promising piezoelectric behavior is attributed to the presence of the morphotropic phase boundary (MPB) in compounds with $x$ = 0.05 - 0.07, similar to PZT.[5,6] However, these solid solutions have a number of disadvantages, such as the low temperature of destruction of the polar ferroelectric state (of the order of 130 – 150 °C) in the solid solutions with compositions from the MPB region of the phase diagram and low values of the piezoelectric modulus $d_{33}$ (160 – 180 $pC/N$).

In majority of publications devoted to investigations of BNT-based solid solutions, these solid solutions were obtained by substitutions in the A-site of the perovskite crystal lattice. In all cases such substitution led to a decrease of the temperature of the ferroelectric − antiferroelectric (FE→AFE) phase transition of the prime BNT. Besides, BNT based solid solutions have large values of the coercivity field which hinders the process of polarization. Due to these factors a replacement of PZT by BNT-based materials does not look possible at present.

It is well known that large values of the piezoelectric parameters in lead-containing compounds have been achieved by means of the *B*-site substitutions. $PbZrO_3 - PbTiO_3$, $Pb(Mg_{1/3}Nb_{2/3})O_3 - PbTiO_3$, and $Pb(Zn_{1/3}Nb_{2/3})O_3 - PbTiO_3$ systems of solid solutions are the perfect examples of this statement.

The purpose of the present paper is to present the results of studies of the influence of ion substitutions into the *B*-site of the crystal lattice in the BNT-based solid solutions. We have investigated the effect of these substitutions on the relative stability of the FE and AFE states and have determined the physical factors that promote the expansion of the temperature range of existence of the FE state.

In PZT solid solutions one of the components ($PbTiO_3$) is a ferroelectric and the other one ($PbZrO_3$) is an antiferroelectric. When the ions substituting titanium have larger sizes the increase of their content leads to the stabilization of the non-piezoelectric AFE phase.[7-10] There are very few reports in the literature (see, for example, Refs 11, 12) in which the titanium ions have been substituted by zirconium in the BNT-based solid solutions. Even less publications are devoted to another *B*-site substitutions.[13-15] However, it has been expected that in the case when the substituting ions are larger then $T^{+4}$-ion, the AFE state is also stabilized.

On the other hand, the substitutions of the smaller ions in the *B*-site of the crystal lattice can stabilize the FE state. Therefore, for investigation of the influence of substitution of titanium we have chosen $[(Na_{0.5}Bi_{0.5})_{0.8}Ba_{0.2}]TiO_3$ as a base solid solution. The location of this solid solution in the phase diagram of the $[(Na_{0.5}Bi_{0.5})]TiO_3 - BaTiO_3$ system is far from the AFE state stability region of BNT.[16,17]

The traditional method of solid state synthesis has been used to obtain BNT based solid solutions. The starting materials were reagent grade oxides and carbonates of corresponding metals. All these starting materials were mixed in the appropriate stoichiometry (except for $Bi_2O_3$, which was taken in excess of 0.5 Wt %) by ball milling during 12 hours. The mixed powders were calcined at 700 − 1100 °*C* for 2 − 20 hours.

The phase analysis has been carried out by means of DRON-3 X-ray diffractometer in the Bragg-Brentano geometry, using Cu*K*α-radiation (λ = 1.5418 Å), Ni-filter for incident beam, and graphite monochromator in the diffraction beam (the angular range was $20^\circ \leq 2\theta \leq 90^\circ$, the scan step was 0.02°, the accumulation time at each point was 1 s). The intermediate phases were identified by means of the ASTM-library.

Synthesized powders were axially pressed into disks with a diameter of 12 mm and a thickness of 1 mm. After that the disks were sintered at 1150 – 1200 °*C* to manufacture samples for dielectric measurements.

A fire-on silver paste was used for electrodes for dielectric and piezoelectric measurements. Temperature dependencies of dielectric parameters were measured at 1 kHz by QuadTech 7600 LCR meter. *D–E* hysteresis loops were observed by a standard Sawyer-Tower circuit at $10^{-2}$ Hz.

To describe the manufacturing of the $[(Na_{0.5}Bi_{0.5})_{0.80}Ba_{0.20}]TiO_3$-based solid solutions let us consider the $[(Na_{0.5}Bi_{0.5})_{0.80}Ba_{0.20}](Ti_{1-y}Zr_y)O_3$ compounds first. The diagram of formation of phases in the solid solution containing titanium and zirconium ions in the *B*-site of the lattice is given in Fig.1 for a particular composition $(Na_{0.5}Bi_{0.5})_{0.80}Ba_{0.20}](Ti_{0.90}Zr_{0.10})O_3$ ($y$ = 0.10). The formation of required solid solution started at 700 °*C* and its share increased when the annealing temperature was elevated. Synthesis of the solid solution is a multi-step process which was accompanied by formation of a number of intermediate phases such as $Bi_{12}TiO_{20}$, $Bi_4Ti_3O_{12}$, $Na_2Ti_6O_{13}$, $Na_2Ti_3O_7$, $Bi_2Ti_2O_7$, $Ba_2TiO_4$, $BaTi_4O_9$, $NaBiTi_6O_{14}$, and $BaZrO_3$ (see Fig.1). However, when the sintering temperature was risen to 1000 – 1100 °*C* a single phase solid solution was obtained.

The sequence of phases formed during the synthesis process was more complicated when the content of zirconium increased to 30 %. New intermediate phases were detected. However, the value of the sintering temperature for the final phase of the solid state synthesis remained 1100 °*C*. A decomposition of solid solutions was not observed with further increase of the sintering temperature up to 1200 – 1250 °*C*.

The optimal sintering temperature in the process of manufacturing of compounds containing Zr or Sn but without Ba was in the range of 900 – 950 °*C*, but the final products were not a single phase solid solution. The increase of the sintering temperature led to the decomposition of solid solutions.

To conclude the discussion of the phase formation in BNT-based solid solutions containing zirconium, we have to discuss the decomposition of solid solutions at high temperatures. The decomposition took place in the solid solutions obtained on the basis of $(Na_{0.5}Bi_{0.5})ZrO_3$ and it did not occur in the solid solutions containing titanium, for example, in $[(Na_{0.5}Bi_{0.5})_{0.80}Ba_{0.20}](Ti_{1-y}Zr_y)O_3$. In the first case, the synthesis of the solid solution did not end with 100% yield of the finished product as well. It should be also emphasized that the introduction of barium into the $(Na_{0,5}Bi_{0,5})ZrO_3$ solid solution reduced the degree of the solid solution decomposition at high temperatures. The decomposition of solid solutions was accompanied by volatilization of sodium oxide.

The transition from $(Na_{0.5}Bi_{0.5})TiO_3$ to $(Na_{0.5}Bi_{0.5})ZrO_3$ was accompanied by an increase in the size of the crystal lattice due to the significant difference in the ionic radii of $Ti^{4+}$ and $Zr^{4+}$ (0.605 and 0.72 Å, respectively, see Ref. 18). One must take into account two major factors. The ionic radius of $Na^+$ is less than that of $Bi^{3+}$ in the 12-coordination position of the perovskite crystal structure. In the case of sodium only one electron participates in the formation of bonds in the crystal lattice, while there are three electrons available for bonding in the case of bismuth. Due to these reasons the increase of interionic distance led to more rapid decrease in the bonding energy for sodium in comparison with bismuth. Therefore, the sodium volatilization and the solid solution decomposition at high temperatures occurred in the case of zirconium containing solid solutions.

In $[(Na_{0.5}Bi_{0.5})_{0.80}Ba_{0.20}](Ti_{1-y}Zr_y)O_3$ solid solutions with zirconium content of 30% the change in the lattice parameters is much less, thus the sodium is coupled to crystal structure stronger than in previous case, and the high-temperature decomposition of solid solution did not take place. As showed above, the formation of the Ba-containing solid solution occurred at higher temperatures than in the absence of zirconium, and the solid state synthesis process was more complicated.

Thus, the synthesis of the $[(Na_{0.5}Bi_{0.5})_{0.80}Ba_{0.20}](Ti_{1-y}Zr_y)O_3$ system of solid solutions is a multistage process which accompanied by formation of different intermediate phases (depending on the composition of the solid solution): $Bi_{12}TiO_{20}$, $Bi_4Ti_3O_{12}$, $Na_2Ti_6O_{13}$, $Na_2Ti_3O_7$, $Bi_2Ti_2O_7$, $Ba_2TiO_4$, $BaBi_4Ti_4O_{15}$, $BaTi_4O_9$, $NaBiTi_6O_{14}$, and $BaZrO_3$. Barium zirconate $BaZrO_3$ was detected in the reaction products after the annealing at 700 °*C* and was preserved in the reaction mixture up to a temperature of 850 °*C*. In the absence of titanium the high-temperature annealing led to the degradation of the synthesized solid solutions. Barium activates the synthesis process, the onset temperature of the solid-state reaction in the $[(Na_{0.5}Bi_{0.5})_{1-x}Ba_x]ZrO_3$ system is reduced compared to the $(Na_{0.5}Bi_{0.5})ZrO_3$. High temperature annealing led to the decomposition of solid solutions, if *B*-sites contained Zr ions only. In solid solution containing titanium (from 100 to 70 %), the decomposition during the high-temperature annealing was not observed.

Ceramic samples for further studies were obtained by sintering of powders at 1200 °*C* for 6 hours. X-ray diffraction patterns for the $[(Na_{0.5}Bi_{0.5})_{0.80}Ba_{0.20}](Ti_{1-y}Zr_y)O_3$ system of solid solutions with $0.00 \leq y \leq 0.30$ are shown in Fig. 2. All samples were single-phase.

The synthesis process of the $[(Na_{0,5}Bi_{0,5})_{0.80}Ba_{0.20}]TiO_3$ based solid solutions with substitution of the $Sn^{4+}$ ion and $(In_{0.5}Nb_{0.5})^{4+}$ and $(Fe_{0.5}Nb_{0.5})^{4+}$ complexes for titanium is simpler than in the case of the $Zr^{4+}$ substitution for Ti. This is caused by the smaller ion radii of the substituting ions and complexes in comparison with the ion radius of $Zr^{4+}$. In the case when solid solutions with $Zr^{4+}$ substituting $Ti^{4+}$ were sintered in the temperature interval 700 – 900 °*C* the intermediated phases appeared, but their number was smaller. In all cases the complete formation of single-phase solid solutions occurred at sintering in a temperature range of 1000-1100 °*C*.

Let us now discuss the influence of ions substitution on the phase transformations. Fig. 3*a* shows the temperature dependencies of the real part of the dielectric constant, $\varepsilon'$**,** obtained on ceramic samples of the $[(Na_{0.5}Bi_{0.5})_{0.80}Ba_{0.20}](Ti_{1-y}Zr_y)O_3$ solid solutions with different zirconium content. Two anomalies are clearly observed. The first anomaly (the low temperature one) corresponds to the FE→AFE phase transition and the second anomaly corresponds to the transition from the AFE (the non-polar phase) to the paraelectric phase (the attribution of phases is made according to Refs. 3, 19). The low temperature phase transition is accompanied by an anomaly in the temperature dependence of the imaginary part of the dielectric constant, $\varepsilon''$**,** (shown in Fig. 3*b*). The temperature, at which the low temperature phase transition takes place decreases significantly when the content of zirconium in solid solution increases. The effect of the zirconium content on the position of the high temperature phase transition was weak.

Our studies of the crystal structure showed that an increase in the content of zirconium led to a linear increase in the size of the unit cell. The dependence of the crystal lattice parameter (calculated in pseudocubic approximation) on zirconium content is given in Fig.4. The dependence is linear with a good level of accuracy.

Similar results were received for solid solutions obtained by the substitution of tin for titanium. The growth of the tin content led to a decrease of the temperature of the transition between the FE and AFE states as well as to a linear increase in the size of the unit cell (sees Fig.4). Since the size of the tin ion is smaller than the size of zirconium one, the rate of increase of the lattice parameter in the case of substitution of titanium by tin was smaller. The temperature of the phase transition between the two dipole-ordered states (FE and AFE) decreased when the tin content increased (see Fig.5).

The influence of the zirconium and tin ions substitutions on the relative stability of the FE and AFE states can be explained based on the size effect, similarly to the explanation of stability of the same phases in the lead zirconate−titanate based solid solutions (PZT). The tolerance factor parameter $t=\left(R_{\mathrm{A}}+R_{\mathrm{O}}\right)/\sqrt{2}\left(R_{\mathrm{B}}+R_{\mathrm{O}}\right)$ is used for this purpose. The FE state is a stable state in the PZT at large values of the tolerance factor parameter ($t$ > 0.9090) while for small values of this parameter ($t$ < 0.9080) the AFE is a stable state.[20] An increase in the radius of the ion in the *B*-site leads to a decrease of the tolerance factor parameter. It is this effect manifests itself in the case of the substitution of titanium ions by zirconium (and also by tin) in the $(Na_{0.5}Bi_{0.5})TiO_3$ based solid solutions. The same conclusion follows from our results of investigation of solid solutions with $Ti^{4+}$ ion substituted by the $(In_{0.5}Nb_{0.5})^{4+}$ complex. The average ion radius of this complex is larger than the radius of $Ti^{4+}$ ion. Therefore, the stability of the FE state has been lowered against the stability of the AFE state when the content of $(In_{0.5}Nb_{0.5})^{4+}$ has increased. At the same time the temperature of the FE→AFE phase transition has decreased.

To verify the proposed dimensional mechanism of influence of ionic substitutions on the FE→AFE phase transition in $(Na_{0.5}Bi_{0.5})TiO_3$-based solid solutions, we used the $(Fe_{0.5}Nb_{0.5})$ complex for substitutions of zirconium and titanium ions during the next stage of our studies. The $(Fe_{0.5}Nb_{0.5})$ complex has the same valence as the replaceable Zr and Ti ions. The substitutions were performed as follows: first we manufactured the $[(Na_{0,5}Bi_{0,5})_{0,80}Ba_{0,20}]\{[(Ti_{0,90}(Fe_{0,5}Nb_{0,5})_{0,1})]_{0,90}Zr_{0.10}]\}O_3$ compound substituting $(Fe_{0,5}Nb_{0,5})^{4+}$ for $Ti^{4+}$ ions and then the $[(Na_{0,5}Bi_{0,5})_{0,80}Ba_{0,20}]\{(Ti_{0,90}[Zr_{0.08}(Fe_{0,5}Nb_{0,5})_{0,02}]_{0.10}\}O_3$ compound substituting $(Fe_{0,5}Nb_{0,5})^{4+}$ for $Zr^{4+}$. In the first case, the average ionic radius of the $(Fe_{0.5}Nb_{0.5})$ complex is greater than the radius of the substituted titanium ion, in the second case the average ionic radius of the $(Fe_{0.5}Nb_{0.5})$ complex is smaller than the radius of the substituted zirconium ion. The temperature dependencies of the imaginary part of the dielectric constant for these solid solutions are given in Fig.6. As one can see, in the first case (in full agreement with the proposed model) the stability of the AFE state with respect to the FE state increases, and at the same time the temperature of the FE→AFE transition decreases. In the second case there is the increase of the stability of the FE state and the temperature of the FE→AFE phase transition increases.

All in all the synthesis of the $(Na_{0.5}Bi_{0.5})TiO_3$ -based solid solutions with substitutions of Ti ions is a multi-step process which is accompanied by the formation of a number of different intermediate phases. Single phase solid solutions have been obtained when the sintering temperatures were 1000 °*C* and above.

The increase of content of substituting ions results in a linear variation in the size of the crystal cell. The relative stability of the AFE and FE phases is in the correspondence with the variation of the tolerance factor of solid solution.

The substitutions of titanium by zirconium and tin lead to a decrease of the tolerance factor and to an increase of the stability of the AFE phase relative to the stability of the FE phase. The same results were obtained in the case of substitutions of Ti by the $(In_{0.5}Nb_{0.5})$ and $(Fe_{0.5}Nb_{0.5})$ complexes.

The substitutions of zirconium by the ($Fe_{0.5}Nb_{0.5}$) complex lead to an increase of the tolerance factor to an increase of the stability of the FE phase relative to the stability of the AFE phase and to an increase of the temperature of the FE→AFE phase transition.

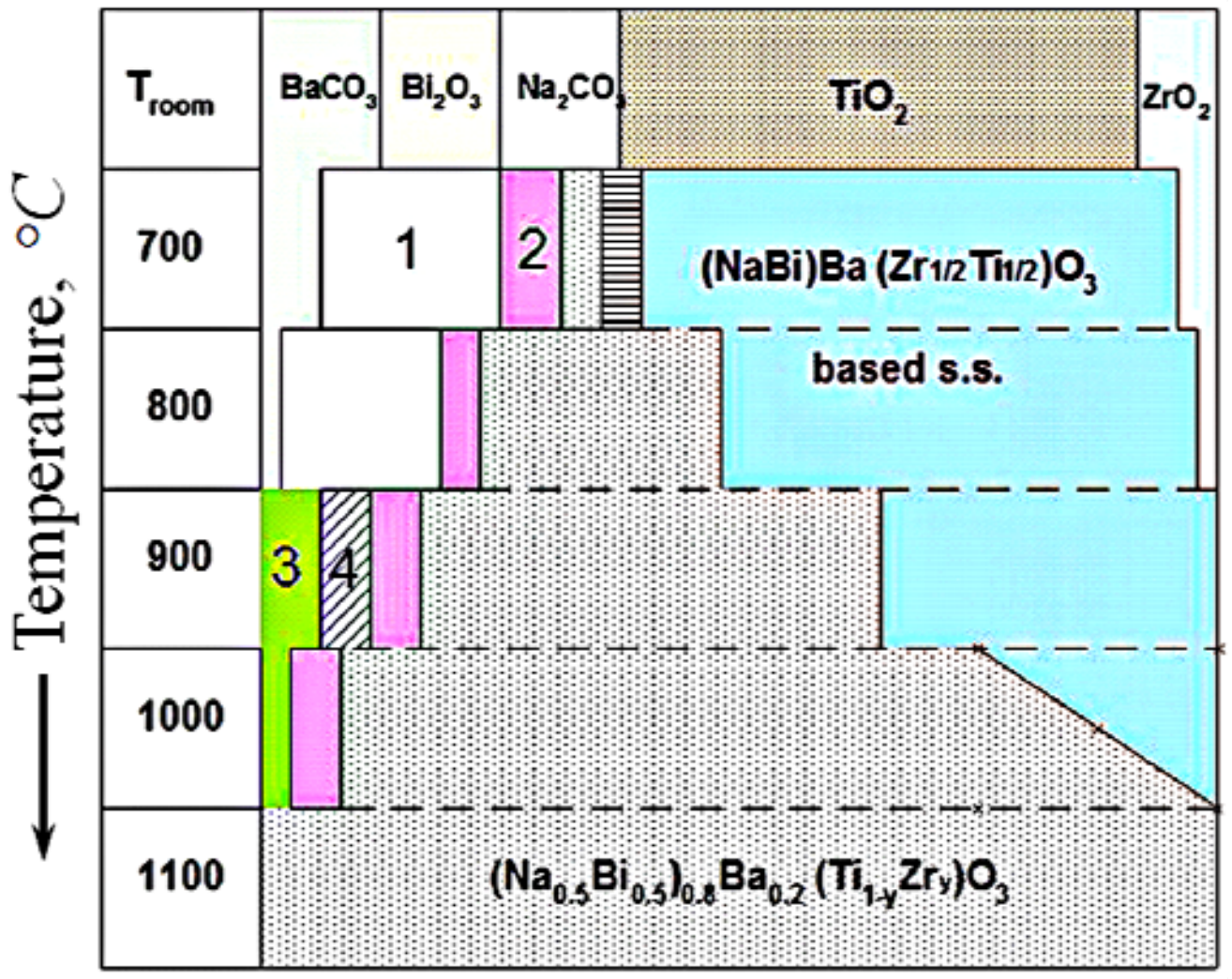

Fig.1. Schematics of transformation of phases during the solid state synthesis of the $[(Na_{0.5}Bi_{0.5})_{0.80}Ba_{0.20}](Ti_{1\text{-}y}Zr_y)O_3$ solid solutions (for $y$ = 10).
Intermediate phases: 1 – $Bi_{12}TiO_{20}$, 2 – $Ba_2Ti_2O_7$, 3 – $Ba_2TiO_4$, 4 – $BaTi_4O_9$.

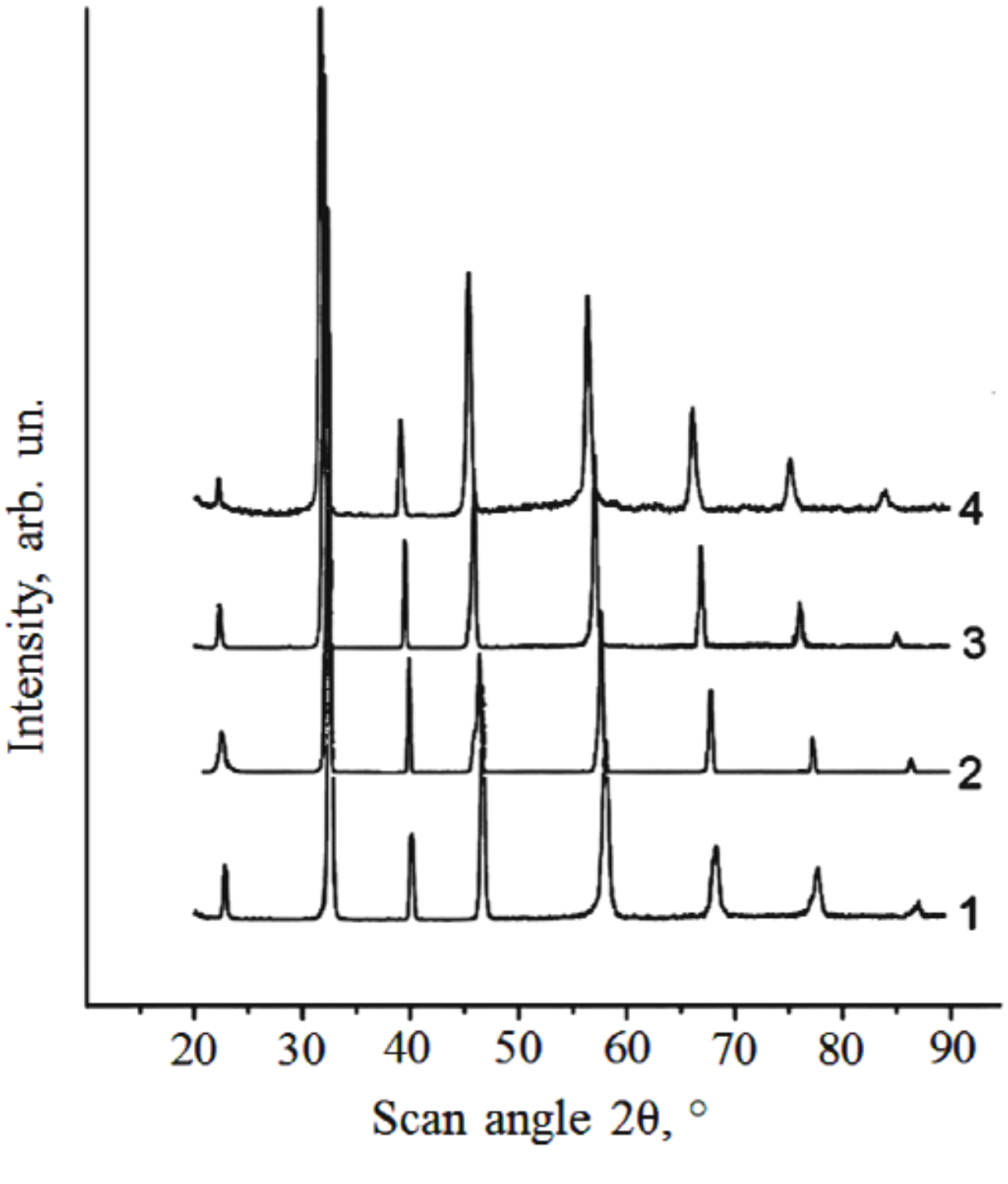

Fig.2. X-ray patterns for the $[(Na_{0.5}Bi_{0.5})_{0.80}Ba_{0.20}]( Ti_{1-y}Zr_y)O_3$ solid solutions.
Zr-content ($y$): 1 – 0.00, 2 – 0.05, 3 – 0.10, 4 – 0.30.

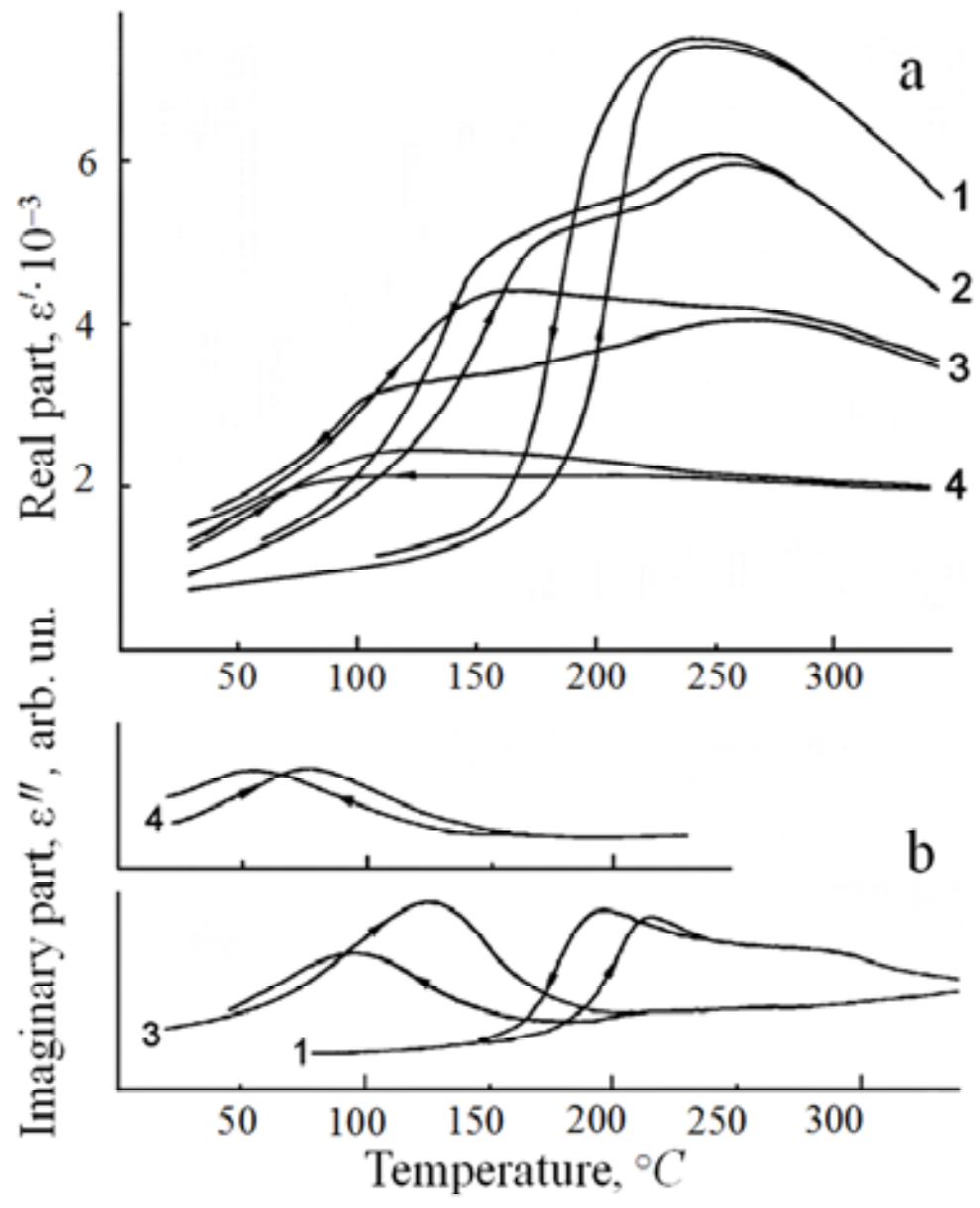

Fig.3. Dependencies of the real (*a*) and imaginary (*b*) parts of dielectric constant on temperature for the $(Na_{0.5}Bi_{0.5})_{0.80}Ba_{0.20}](Ti_{1-y}Zr_y)O_3$ solid solutions.
Zr-content, *y*: 1 – 0.00, 2 – 0.025, 3 – 0.05, 4 – 0.10.

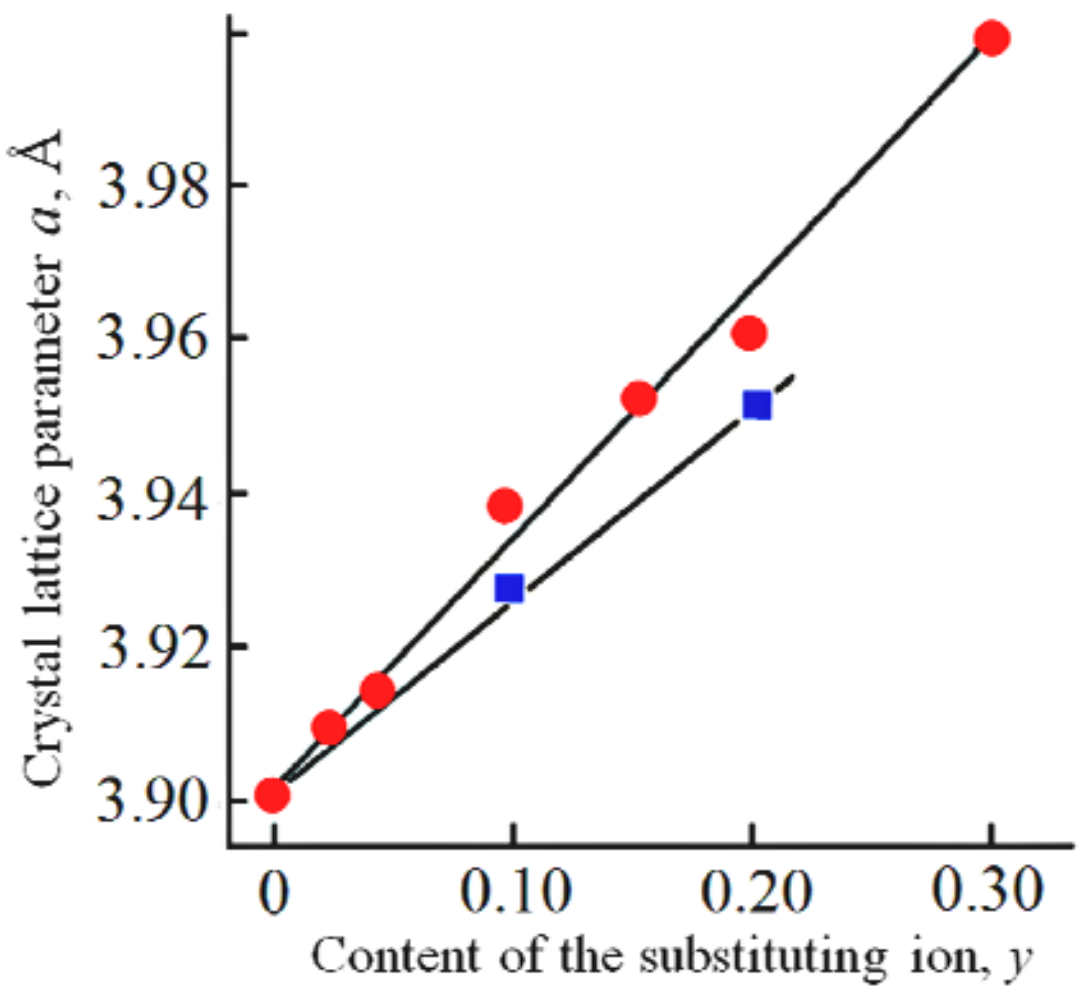

Fig.4. Dependencies of the crystal lattice parameter on Zr and Sn content for the $[(Na_{0.5}Bi_{0.5})_{0.80}Ba_{0.20}](Ti_{1-y}B_y)O_3$ -based solid solution.
Substituting element, B: ● – Zr, ■ – Sn.

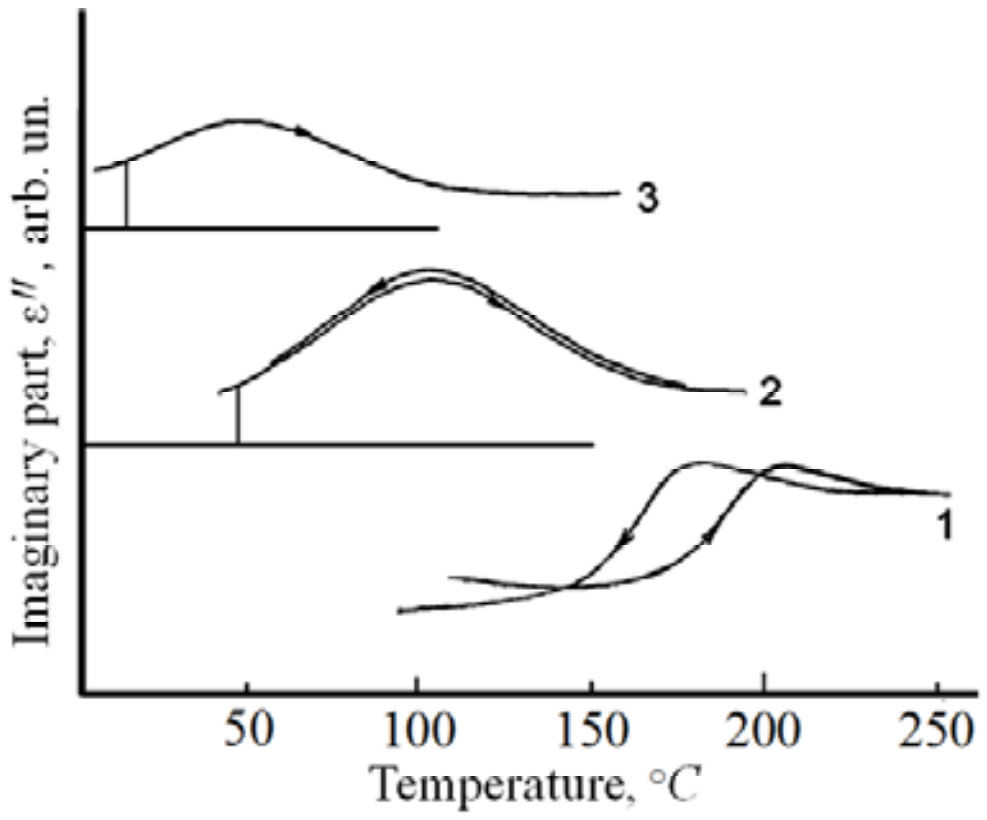

Fig.5. Imaginary part of dielectric constant vs. temperature for the $(Na_{0.5}Bi_{0.5})_{0.80}Ba_{0.20}](Ti_{1-y}Sn_y)O_3$ solid solutions. Sn content, *y*: 1 – 0.00, 2 – 0.10, 3 – 0.20.

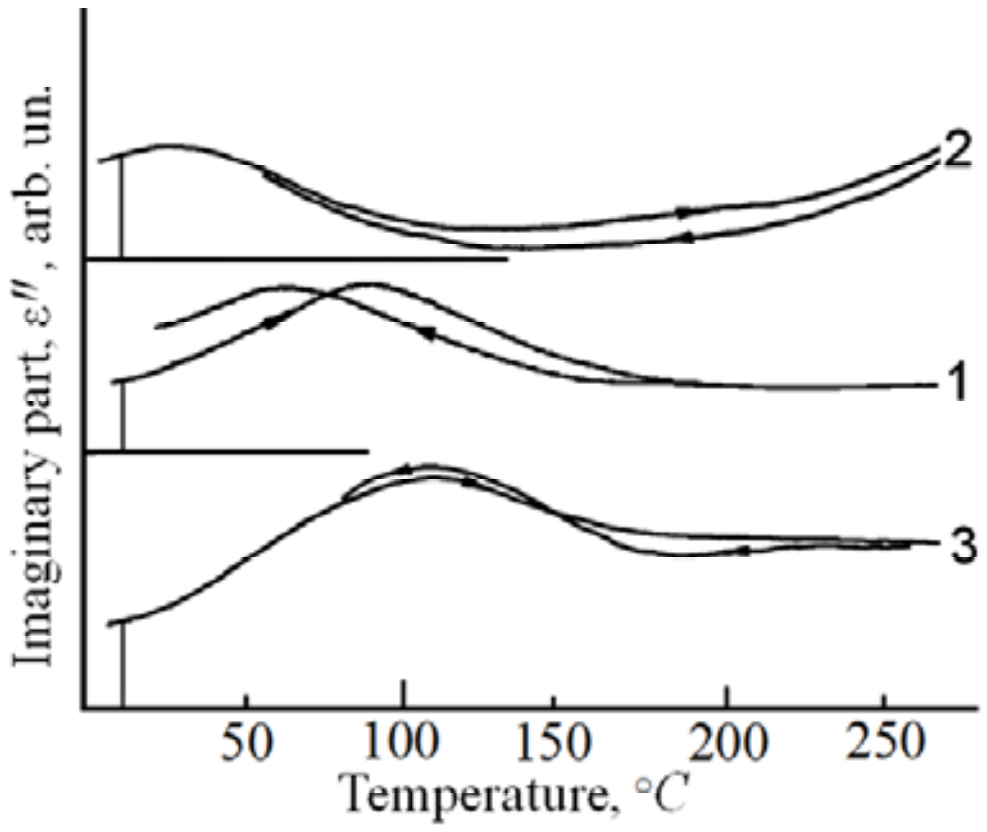

Fig.6. Dependencies of the imaginary part of dielectric constant on temperature for the $(Na_{0.5}Bi_{0.5})_{0.80}Ba_{020}](Ti_{0.9}Zr_{0.10})O_3$ -based solid solutions.
1 – without Ti- and Zr-substitution in the B-site
2 – with substitution of $(Fe_{0.5}Nb_{0.5})$ complex for Ti according to the chemical formula $(Na_{1/2}Bi_{1/2})_{0.80}Ba_{0.210}][\{[Ti_{0.9}(Fe_{1/2}Nb_{1/2})_{0.10}]_{0.90}Zr_{0.10}\}O_3$
3 – with substitution of $(Fe_{0.5}Nb_{0.5})$ complex for Zr according to the chemical formula $[(Na_{1/2}Bi_{1/2})_{0.80}Ba_{0.20}]\{Ti_{0.9}[Zr_{0.08}(Fe_{1/2}Nb_{1/2})_{0.02}]_{0.10}\}O_3$.